\documentclass[showpacs,preprintnumbers,10pt,twocolumn]{revtex4}%
\usepackage{amssymb}
\usepackage{amsfonts}
\usepackage{amsmath}
\usepackage{graphicx}
\usepackage{times}
\usepackage{dcolumn}
\usepackage{bm}
\usepackage{revsymb}
\usepackage{color}%
\usepackage{epstopdf}
\usepackage[normalem]{ulem}

\begin{document}
\title{Experimental and numerical observation of dark and bright discrete solitons in the band-gap of a diatomic--like
electrical lattice}
\author{F. Palmero$^{1}$, L. Q. English$^{2}$, Xuan-Lin Chen$^{3}$, Weilun Li$^{2}$, Jes\'{u}s Cuevas-Maraver$^{4}$, and P. G. Kevrekidis$^{5}$}
\affiliation{$^{1}$Grupo de F\'{\i}sica No Lineal, Departamento de F\'{\i}sica Aplicada I,
Escuela T\'{e}cnica Superior de Ingenier\'{\i}a Inform\'{a}tica, Universidad
de Sevilla, Avda Reina Mercedes s/n, E-41012 Sevilla, Spain}
\affiliation{$^{2}$Department of Physics and Astronomy, Dickinson College, Carlisle,
Pennsylvania, 17013, USA}
\affiliation{$^{3}$ Physics Department, Harbin Institute of Technology, Harbin 150001, Heilongjiang Province, China}
\affiliation{$^{4}$Grupo de F\'{\i}sica No Lineal, Departamento de F\'{\i}sica Aplicada I, Escuela Polit\'{e}cnica Superior, Universidad de Sevilla, Virgen de \'{A}frica 7, 41011 Sevilla, Spain and Instituto de Matem\'{a}ticas de la Universidad de Sevilla (IMUS), Edificio Celestino Mutis, Avda Reina Mercedes s/n, E-41012 Sevilla, Spain}
\affiliation{$^{5}$ Department of Mathematics and Statistics, University of Massachusetts, Amherst, Massachusetts 01003, USA}

\date{\today}

\begin{abstract}
  We observe dark and bright intrinsic localized modes (ILMs) or discrete breathers (DB) experimentally and numerically in a
  diatomic--like electrical lattice. The generation of dark ILMs by driving a
  dissipative lattice with spatially-homogenous amplitude is, to our knowledge,
  unprecedented. In addition, the experimental manifestation of bright
  breathers within the bandgap is also novel in this system. In experimental
  measurements the dark modes appear just below the bottom of the top
  branch in frequency. As the frequency is then lowered further into the
  band-gap, the dark DBs persist,
  until the nonlinear localization pattern reverses and bright DBs appear
  on top of the finite background. Deep into the bandgap, only a single bright structure survives
  in a lattice of 32 nodes. The vicinity of the bottom band also features bright and dark self-localized excitations. These results pave the way for a more
  systematic study of dark breathers and their bifurcations in diatomic--like chains.
\end{abstract}
\maketitle

\section{Introduction} 

The study of discrete breathers was arguably initiated by the works
of~\cite{ST,TKS} on the so-called Fermi-Pasta-Ulam-Tsingou lattices.
It was subsequently significantly propelled forward by the rigorous
proof of their existence in a large class of models starting from
the uncoupled, so-called anti-continuous limit~\cite{MA}.
Since then, there has been a wide variety of systems in which
experimental observations of structures have been reported that are exponentially localized in space and
periodic in time. Among them,
one can note arrays of Josephson junctions \cite{Trias,Ustinov},
mechanical \cite{Lars, Pal16} and magnetic pendula \cite{Russell2},
microcantilevers \cite{Sato}, electrical chains \cite{electric1,Faustino2},
granular crystals, \cite{granular2,granular}, ionic crystals such as
PtCl \cite{Swanson} and antiferromagnets \cite{Sato2}. Moreover, these
types of structures have been offered as plausible explantions of
experimental findings in settings such as the DNA denaturation~\cite{Peyrard},
~$\alpha$-uranium~\cite{Uranium} or NaI~\cite{NaI}; see also the
relevant review of~\cite{flachrev}.

A significant playground for the exploration of discrete breathers, since
early on, has been offered by the setting of electrical
lattices~\cite{remoiss}. Progressively, over the last decade, these
lattices have enjoyed a larger degree of control of their external
drive and lattice configurations (one- vs. two-dimensional, nearest-
vs. beyond nearest-neighbor configurations, etc.)~\cite{electric1,electric2,electric3,electric4,electric5, electric6}.
In these works, a qualitatively and semi-quantitatively accurate model
of the dynamics has also been put forth. Building on this existing
experimental, theoretical and computational background, we now advance
a diatomic-like electrical lattice model.

More specifically, we construct a coupled-resonator circuit where one of the
inductive elements alternates between two values for adjacent nodes,
endowing the lattice with a diatomic structure.
Such electrical lattices have been proposed (and even experimentally
explored) since the early days
of solitonic dynamics; see, e.g.,~\cite{kofane}. Here, however,
we explore some features not previously discussed, to the best
of our knowledge. 
After confirming the
linear two-band spectrum, we search the bandgap for the existence
of nonlinear excitations. Both in the experiment and in the numerical
simulations of the model, we identify breather structures consisting
of density dips in a finite background, namely the so-called
dark breathers (DBs). To the best of our knowledge, such dark breathers
have previously been induced experimentally only in granular
crystals~\cite{granular}, on the basis of earlier
theoretical predictions in Hamiltonian and dissipative lattices, such as~\cite{azucena,guill,chong2}. Nevertheless,
the latter setting only affords control of the driving boundaries,
while no drive or spatio-temporal controls are available within
the chain, as is the case herein. Hence, it can be argued that
the present setting is far more amenable to the controllable
generation and persistence of these states. Dark breathers
are found in the vicinity of the top band and, as we drive
deeper in the spectral gap, the nonlinear states sustain a fundamental
change of character and become bright ones, on top of the finite
background (which previously supported the DBs).  In addition, we have found that below the bottom band the system features dark and bright ILMs structures as well. To the best of our knowledge such features are unprecedented in electrical chains or, for that matter, in experimentally tractable nonlinear dynamical
lattices generally.

Our presentation is structured as follows. First, we present the
experimental setup and corresponding theoretical model. We then
discuss our results for both dark and bright discrete breathers.
Finally, we summarize our findings and present some directions
for future study. 

\section{Experimental and theoretical setup} 

In our electrical lattice, the connected
nodes are RLC resonators comprised of inductors $L_{2}^{(a,b)}$ and varactor diodes providing a voltage-dependent, nonlinear capacitance, $C(V)$.  In order to make this lattice a diatomic system, we have incorporated two different cells, 
$(a)$ and $(b)$, with different values of inductors, $L_2^{(a)}$ and $L_2^{(b)}$, located on alternate sites. Thus, the unit cell is the node-pair, and each node of one type is surrounded by two nodes of the other type. Neighboring nodes are coupled inductively via inductors, $L_1$, and driven by a sinusoidal signal via a resistor, $R$. We also connect the last circuit cell to the first, thus implementing periodic boundary
conditions.

\begin{figure}[h]
\includegraphics[width=0.45\textwidth]{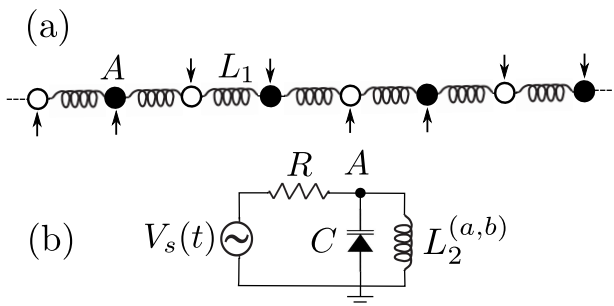}
\caption{(a) Schematic circuit
diagrams of the electrical lattice, where the blank points represents 
circuit cells of type $(a)$ and black points
represent circuit cells of type $(b)$. Black arrows represent driving 
and its phase. (b) Each cell is connected to a
periodic voltage source $V_{s}(t)$ via a resistor $R$, and grounded. Each
point $A$ of an elemental circuit is connected via inductors $L_{1}$ to the
corresponding points $A$ of neighboring cell. Voltages are
monitored at point $A$.}%
\label{lattice}%
\end{figure}

The sinusoidal driver is spatially homogenous in amplitude, but staggered in space from one unit cell to the next. This  means that we instituted a phase shift of $\pi$ between two consecutive nodes of the same type, so that the zone boundary plane-wave modes could be excited. (The two sites within a given unit cell were driven in phase.) We denote the driver amplitude by $V_d$, and its angular frequency by $\omega_d$. A sketch of the system is shown in Fig. \ref{lattice}.

Furthermore, the concrete parameter values of the system
(comprised of inductors, resistors and capacitors) are
$L_1=0.68$ mH, $L_2^{(a)}= 102.7$ $\mu$H, $L_2^{(b)}= 970$ $\mu$H, $R=10$ k$\Omega$ and a NTE 618 varactor, for which the capacitance at zero voltage is $C_0=770$ pF. The number of nodes of each type was $N^{(a)}=N^{(b)}$ = 16, equaling 16 unit cells for a total of 32 nodes. Experimentally, node voltages $V_n$ at points $A$ were measured
at a rate of 2.5 MHz using a multichannel analog-to-digital converter. 

It has been shown that the basic dynamics of this electrical
network can be qualitatively described by a simple model where the introduction of an amplitude-dependent phenomenological
resistor, $R_l$, in parallel to the cell capacitance, is enough to reproduce experimentally observed
features quite well \cite{electric1}. Accordingly, the dimensionless equations of motion read
\begin{eqnarray}
\label{govern}
\frac{d i_n}{d \tau}=\frac{L_2^{(a)}}{L_1}(v_{n+1}+v_{n-1}-2v_n)-\frac{L_2^{(a)}}{L_2^{(n)}} v_n \\ \nonumber
\frac{d v_n}{d \tau}=\frac{1}{c(v_n)}\left[i-i_{d}+\frac{\cos(\Omega \tau)}{\omega_0 R C_0}-\frac{v_n}{\omega_0 R_e C_0}\right],
\end{eqnarray}
where $L_2^{(n)}=L_2^{(a)}$ if site $(n)$ corresponds to nodes $(a)$, or $L_2^{(n)}=L_2^{(b)}$ if it corresponds to nodes $(b)$. The following dimensionless variables have been used: $\tau = \omega_0 t$; $i_n = (I_v - I_2)/(\omega_0 C_0 V_d)$, where $I_v$ is the full current through the unit cell and $I_2$ is the current 
through the inductor $L_2^{(n)}$, both corresponding to cell $n$; $v_n = V_n/V_d$. Here, $V_d$ is the driver amplitude, and $V_n$ is 
the measured voltage at node $n$; $\Omega = \omega_d/\omega_0$, $\omega_0 = 1/\sqrt{L_2^{(a)}C_0}$; $i_d=I_D/(\omega_0 C_0 V_d )$, where $I_D$ is the current through the varactor diode; $c = C(V )/C_0$, and $C(V )$ is the nonlinear capacitance of the diode. $R_e$ is the equivalent resistance so $1/R_e=1/R+1/R_l$.

\section{Results and Discussion}

\begin{figure}[h]
\includegraphics[width=0.55\textwidth]{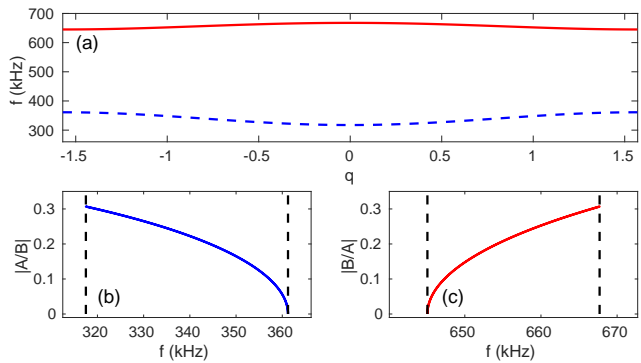}
\caption{(a) Electrical line linear mode frequencies $f$ as function of the wavevector $q=n \pi/N$, where $n=-N/2 \ldots N/2$ and $N=N^{(a)}+N^{(b)}$.
      A solid line is used to denote the top and a dashed to denote
    the bottom band. Figures (b) and (c) show the ratios between different linear mode amplitudes in the bands, where $A$ correspond to oscillation amplitudes of cells of type $(a)$ and $B$ to cells of type $(b)$.}%
\label{band}
\end{figure}

\begin{figure}[h]
\includegraphics[width=0.5\textwidth]{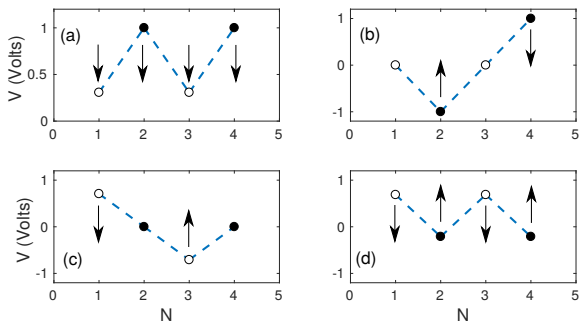}
\caption{Different linear modes corresponding to the top and the bottom of the bands shown in Fig. \ref{band} . (a) Uniform linear mode corresponding to the bottom of the lower band. (b) Linear mode corresponding to the top of the lower band, where the $L_2^{(b)}$ sub-lattice oscillates out-of-phase and the other sub-lattice is at rest. (c) Linear mode corresponding to the bottom of the upper band, where the $L_2^{(a)}$ sub-lattice oscillates out-of-phase, and the other sub-lattice is at rest. (d) Linear mode corresponding to the top of the upper band, where all neighboring cells oscillate out-of phase.
}%
\label{phonon}
\end{figure}

In the limit of no driving and for small-amplitude oscillations  
%the dispersion relation is given by
%\begin{equation}
%\omega^2=\frac{(a+b)\pm \sqrt{(a+b)^2-4(ab-d\cos^2 q)}}{2},
%\end{equation}
%where $a=1/C_0L_2^{(a)}+2/C_0L_1$, $b=1/C_0L_2^{(b)}+2/C_0L_1$, $d=4/C_0^2L_1^2$ and $q \in [0,  \pi]$.
the linear modes possess frequencies grouped  into  two bands, as
expected on the basis of the diatomic structure of the lattice.
The top (optic-like) and bottom (acoustic-like) bands of the linear spectrum are found to
lie within the intervals $[314-356]$ and $[645-666]$ kHz, respectively. A plot of the 
theoretical dispersion bands - obtained from linearizing Eq.~(\ref{govern}) - is shown in Fig. \ref{band}, and the linear modes corresponding to top and bottom of the two bands are 
shown in Fig.~\ref{phonon}. Using the spatial signature of the driving outlined before, we are able to experimentally excite and pump the linear mode of Fig.~\ref{phonon}(c) and (b), and we proceed to explore the dynamics of such
a mode for frequencies within the bandgap (i.e., in the regime of nonlinear,
breather-like excitations~\cite{flachrev}). 

\subsection{Dark gap breathers}
The existence of bright ILMs below the {lower} linear
dispersion band has been well established in such systems, driven either directly or with a subharmonic driver \cite{electric1, electric2, electric3, electric4, electric5}. In this work, we focus on the existence and properties of discrete breathers in the band-gap, where we have, remarkably, found both dark and
bright breathers (with non-vanishing background), depending on the precise driver frequency within the gap. 

\begin{figure}[h]
\includegraphics[width=0.27\textwidth]{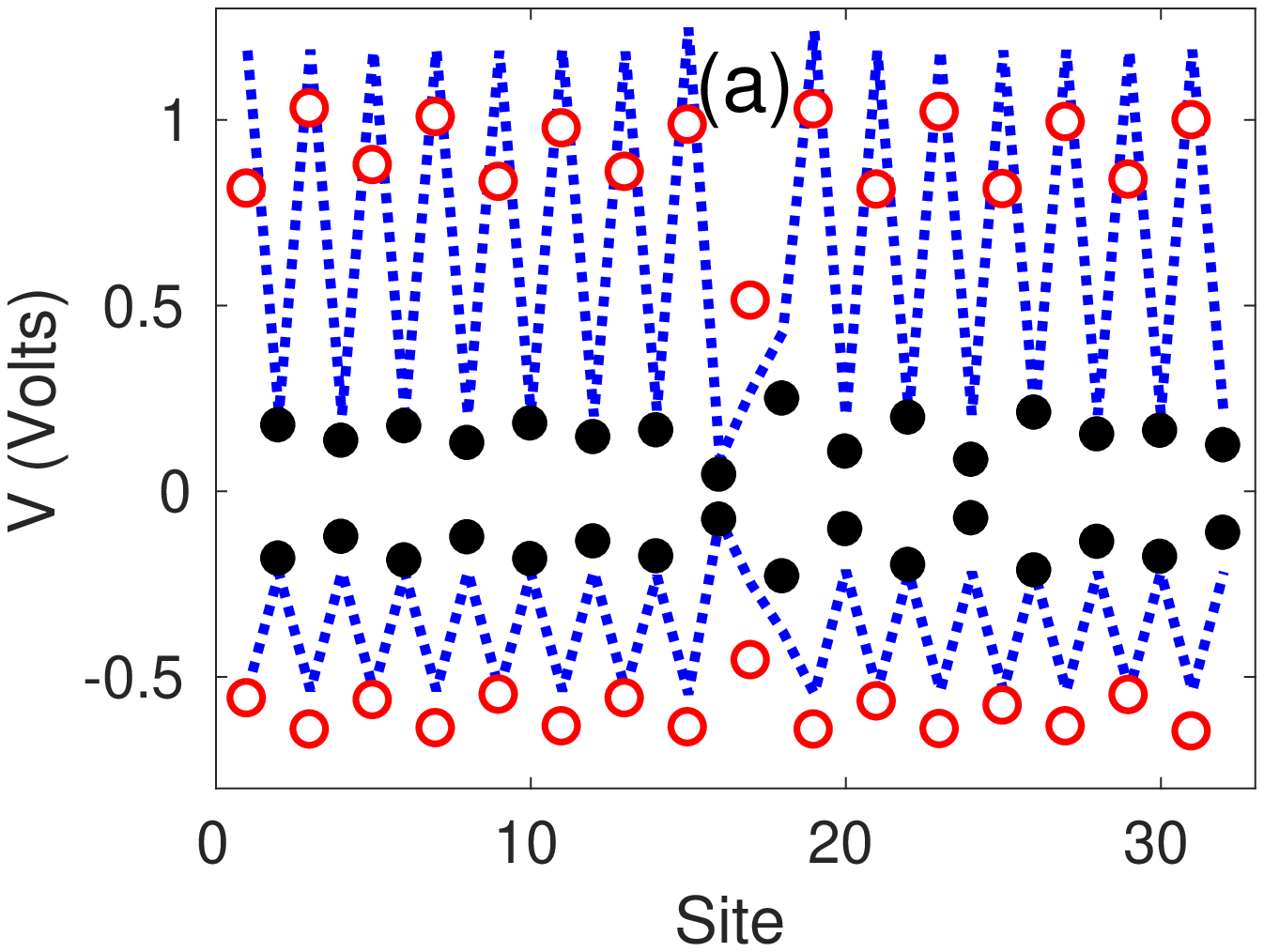} 
\includegraphics[width=0.20\textwidth]{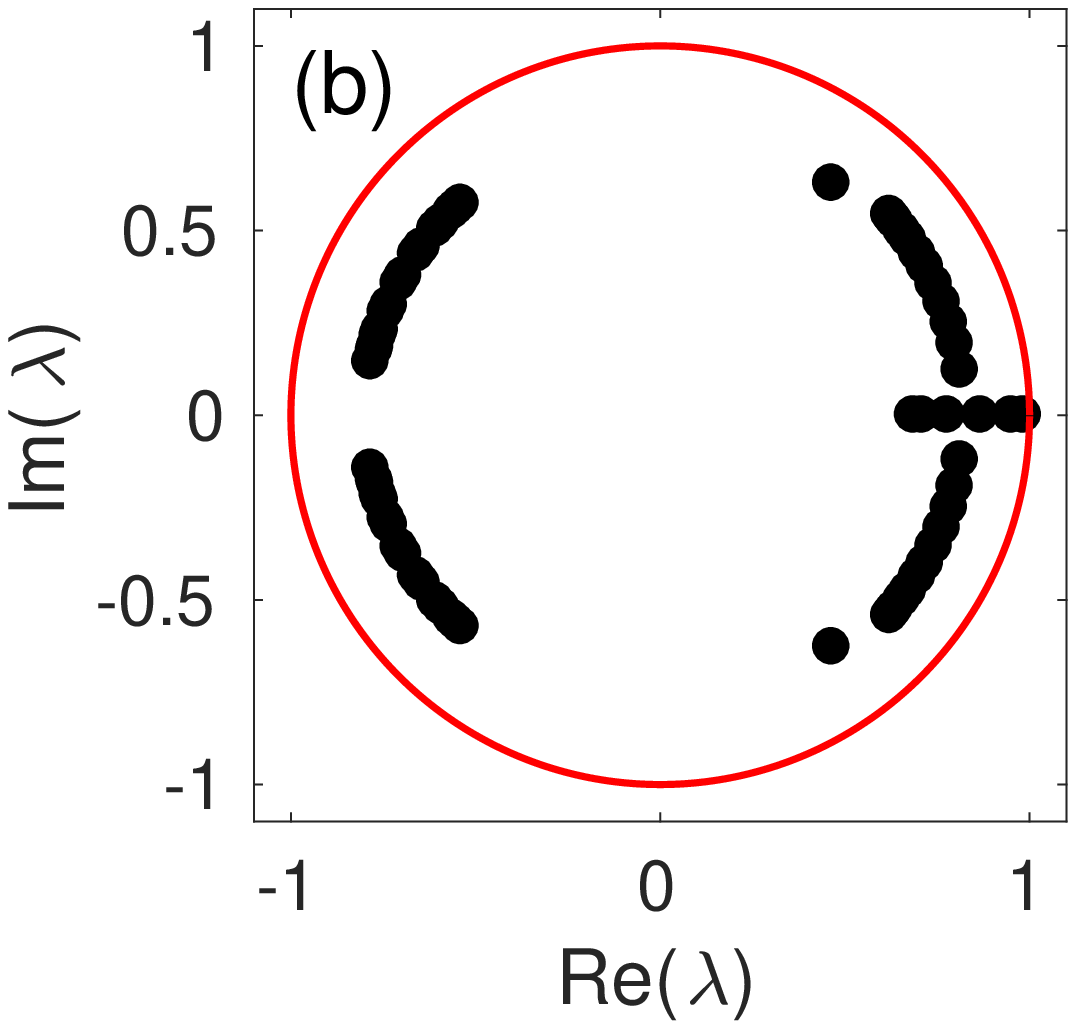}
\caption{(a) Numerical (dotted line) and experimental (circles) dark breather profile (maximum and minimum amplitude) corresponding to $V_d=3.5$ Volts and $f=571$ kHz. Blank points represent circuit cells of type $(a)$ and black points represent circuit cells of type $(b)$. (b) Floquet multiplier numerical linearization spectrum  confirming that, since all multipliers are inside the unit circle, the stability of the solution}%
\label{dark_1}
\end{figure}

\begin{figure}[h]
	\includegraphics[width=0.23\textwidth]{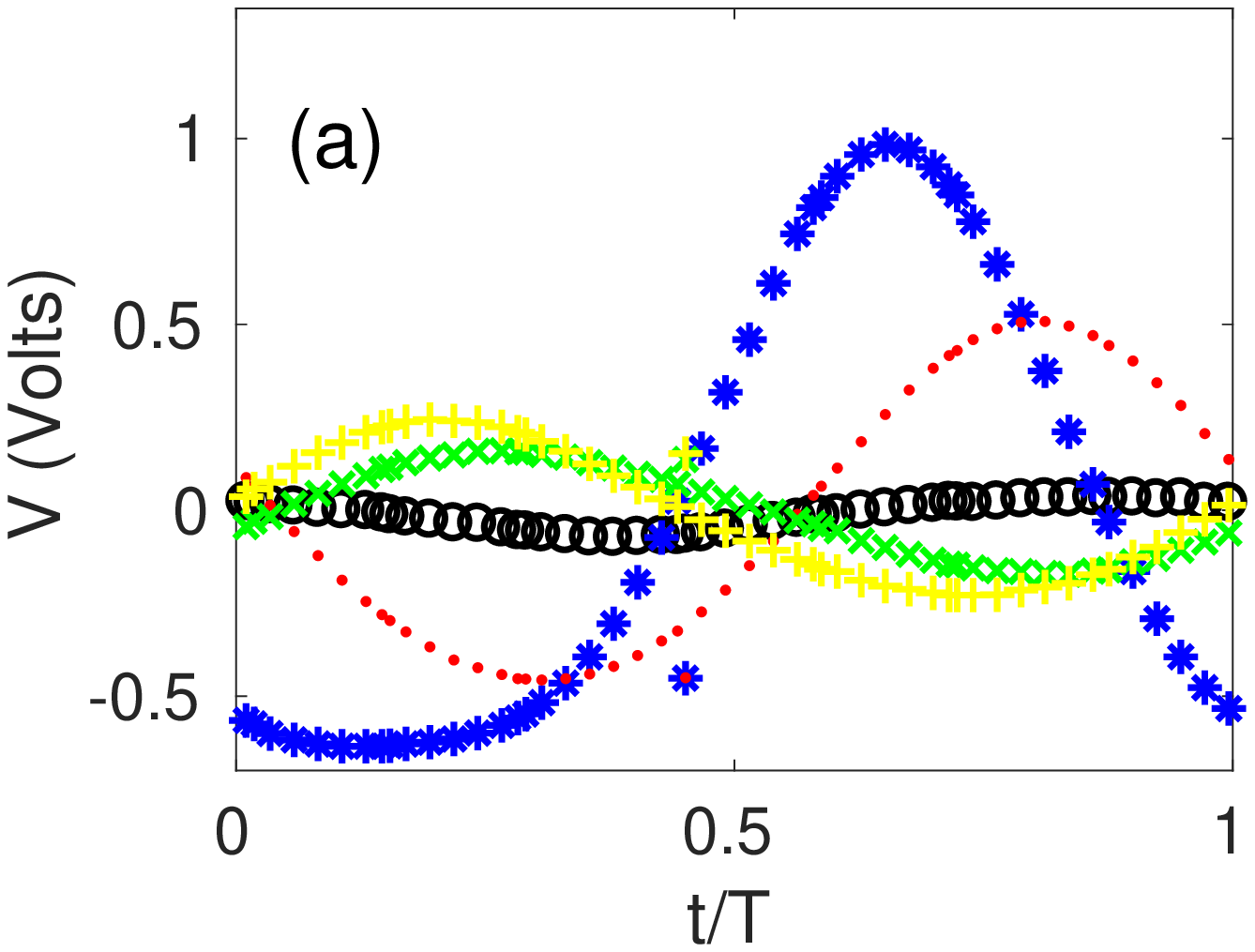} 
	\includegraphics[width=0.23\textwidth]{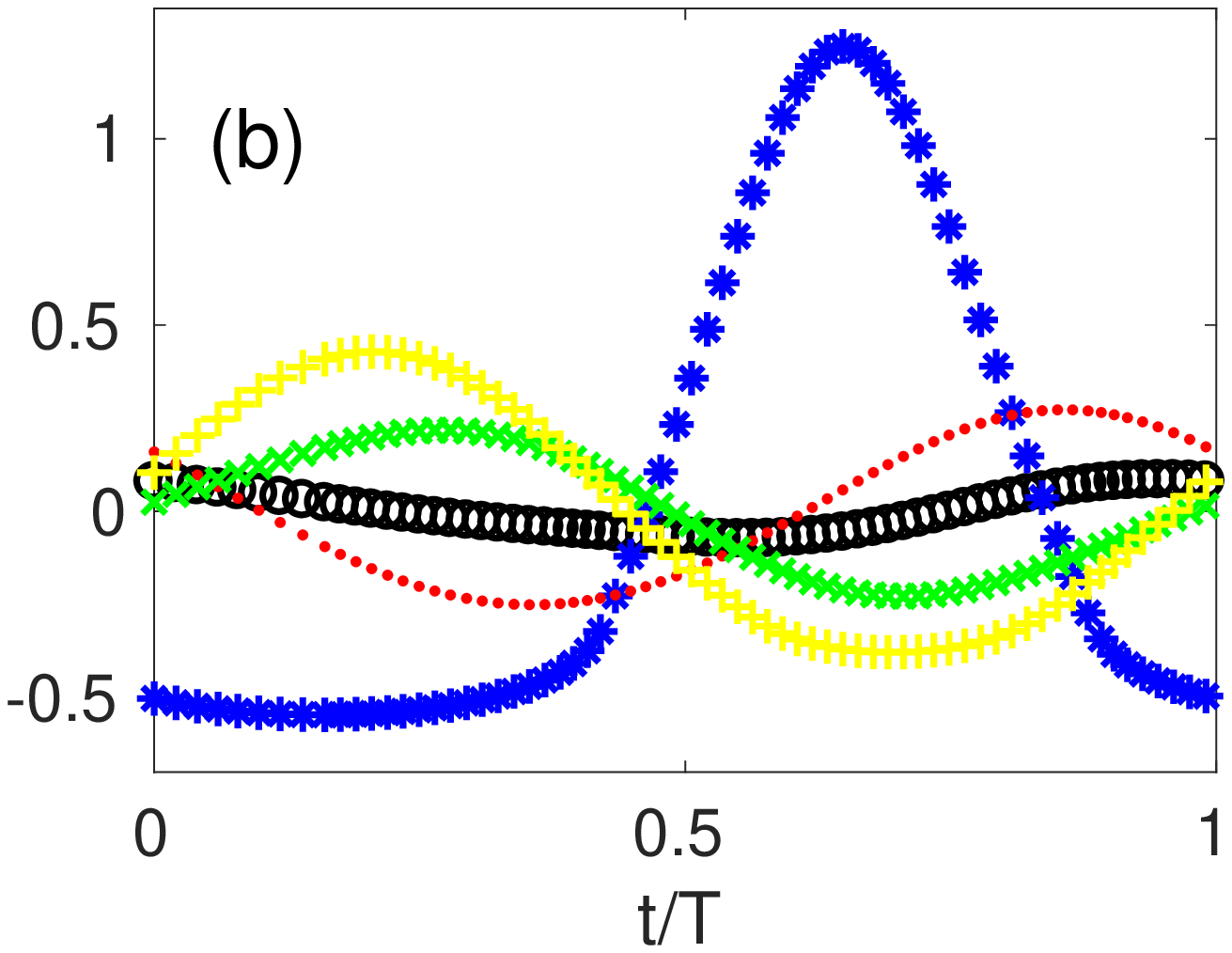}
	\caption{(a) Experimental and (b) numerical voltage oscillations of the dark breather corresponding to $V_d=3.5$ Volts and $f=571$ kHz: site 16 (black circles), first neighbors sites 15 (blue,diamond) and 17 (red
          dots), and second neighbors: sites 14 (green crosses) and 18
          (yellow pluses).}%
	\label{dark_1_t}
\end{figure}

Figure \ref{dark_1} shows the experimental result for driving at a frequency of 571 kHz, about 25\% into the gap from above, and 74 kHz below the lower band-edge of the top band (the zone-boundary mode is at 645 kHz). The amplitude is $V_d=3.5$ Volts. The lattice profile of maximum and minimum node voltages in Fig.~\ref{dark_1} clearly indicates the presence of a dark localized mode located around nodes 16 and 17. The term ``dark'' is used to describe a localized mode for which the amplitude of oscillation is large and constant in the lattice except at the ILM center and in its close vicinity (where it is
near vanishing).
Such structures, although theoretically proposed early on in~\cite{guill},
were only identified quite recently in a materials system (a granular
crystal) in the work of~\cite{granular2}. There, they were produced as
a result of the destructive interference of waves emanating from
the boundary, while here they are robustly created by the drive within the
bandgap, as evidenced in the computational stability analysis of
Fig.~\ref{dark_1}.
Figure \ref{dark_1_t} shows the experimental time series at selected nodes. On one sublattice, node 17 (red dots) is suppressed relative to node 15 (blue
diamonds), and on the other sublattice, node 16 (black, open circle) is suppressed relative to nodes 14 (green crosses) and 18 (yellow pluses).  Experimentally (left panel), we see that the amplitude on the large-oscillation sublattice is reduced at the dark ILM center to about half  the value in the wings. Numerically (left panel) the reduction is
similarly (or even more) pronounced for these particular driving conditions. 

In other experimental runs, the center of the dark ILM forms at different locations within the lattice, so this is not an impurity effect and as expected
in a longer (nearly) homogeneous lattice the DB enjoys the discrete shift
invariance of the diatomic chain. Nevertheless, we also observe that the the location can be predetermined (i.e., designed) by placing a temporary impurity
at the chosen site before turning on the driver. The sign of the impurity necessary to seed a dark ILM is opposite to the one needed for bright ILMs. Numerically, we find stable one--peak dark breathers existing in the range of [569.5-600] kHz for a drive amplitude of $V_d=3.5$ Volts. It remains
an important, albeit numerically rather demanding, question as to whether the dark breather can be found in computations throughout the
system's bandgap.

\begin{figure}[h]
	\includegraphics[width=0.27\textwidth]{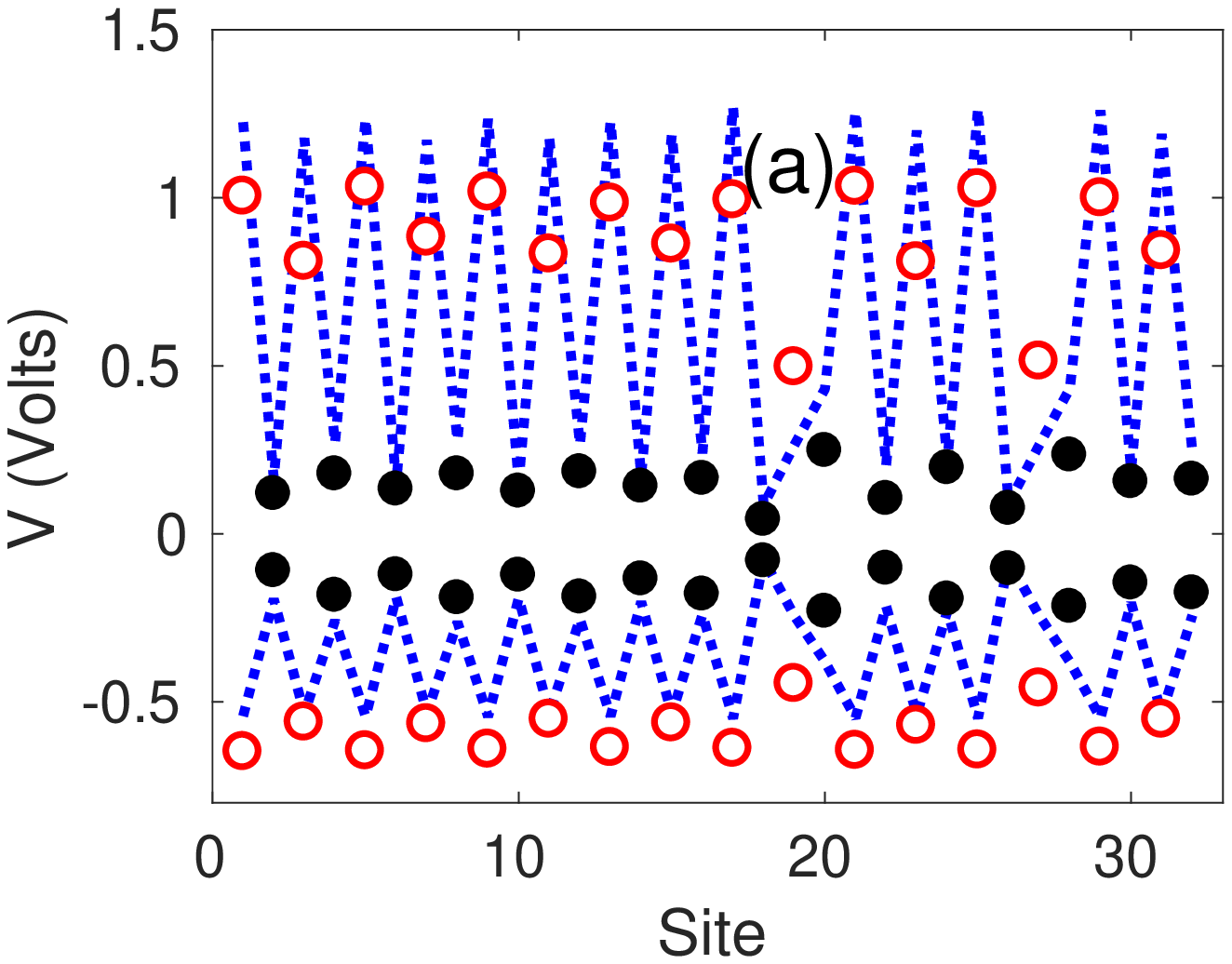}
	\includegraphics[width=0.20\textwidth]{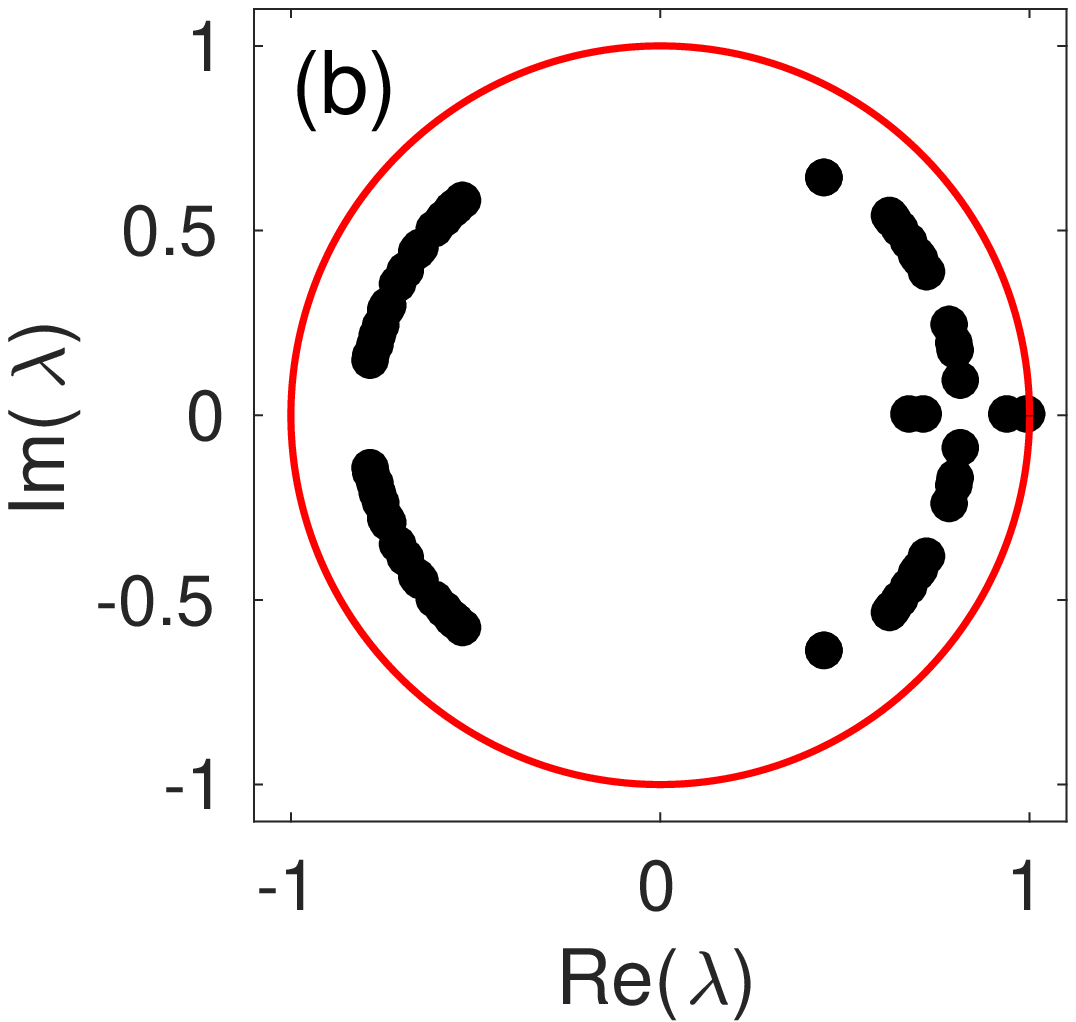}
	\caption{(a) Numerical (dotted line) and experimental (circles) two--peak dark breather profile (maximum and minimum amplitude) corresponding to $V_d=3.5$ Volts and $f=570$ kHz. Blank points represent circuit cells of type $(a)$ and black points represent circuit cells of type $(b)$. (b) Floquet multiplier numerical linearization spectrum confirming the stability of the solution since all multipliers are inside the unit circle.}%
	\label{dark_2}
\end{figure}

\begin{figure}[ht]
	\includegraphics[width=0.27\textwidth]{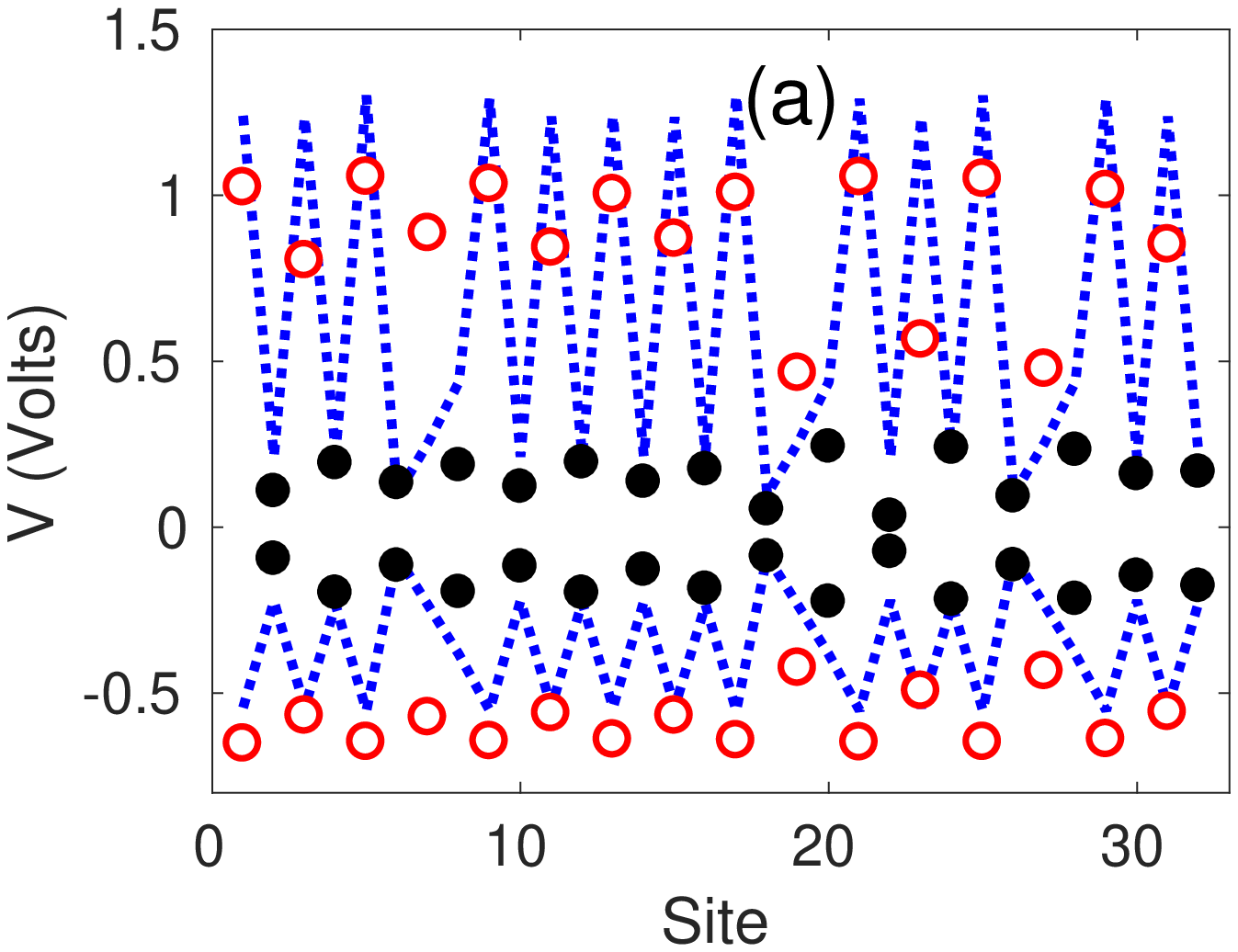}
	\includegraphics[width=0.20\textwidth]{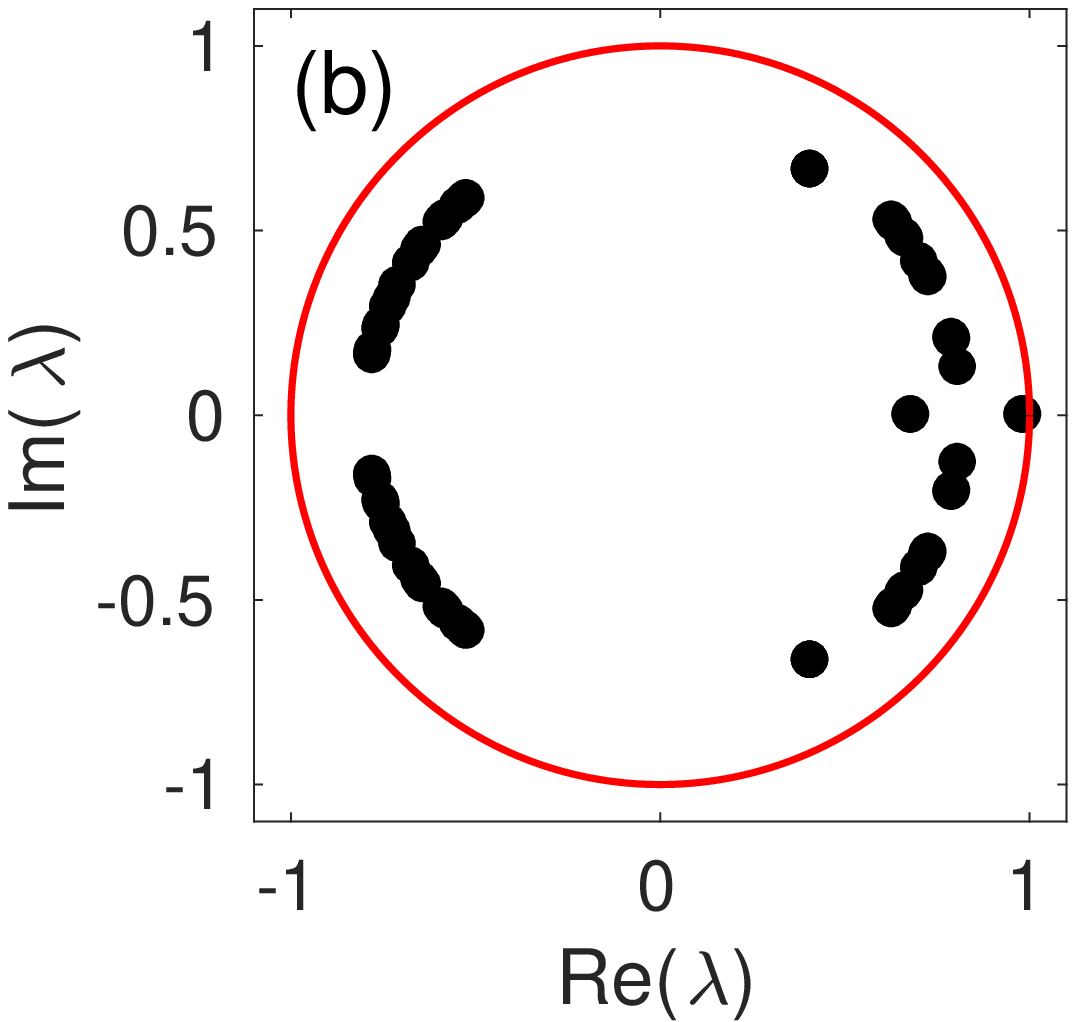}
	\caption{(a) Numerical (dotted line) and experimental (circles) three--peak dark breather profile (maximum and minimum amplitude) corresponding to $V_d=3.5$ Volts and $f=567$ kHz. Blank points represent circuit cells of type $(a)$ and black points represent circuit cells of type $(b)$. (b) Floquet multiplier numerical linearization spectrum  confirming the stability of the solution since all multipliers are inside the unit circle.}%
	\label{dark_3}
\end{figure}

Decreasing the driving frequency further into the bandgap, a pattern of multi-peak dark ILMs emerges, as shown in Fig. \ref{dark_3} . The number of dark ILMs in the lattice is very sensitive to the precise driver frequency at constant amplitude. In general, decreasing the driving frequency produces multi-peak structures and numerical solutions show overall good agreement with experimental results. However, numerics and experiments differ as to the precise localization centers, and in experiments they are found to be spaced somewhat more closely than in the
numerical results.
Nevertheless, the numerical results strongly suggest that such coherent
structures should be dynamically robust, in line with the corresponding
experimental observations.
As also seen in previous studies, where the ILM is
generated by a uniform driver, it can be sensitive to small lattice impurities
which are, unfortunately, hard to completely eliminate
in a manufactured lattice.

\subsection{Bright gap breathers}

\begin{figure}[h]
\includegraphics[width=0.27\textwidth]{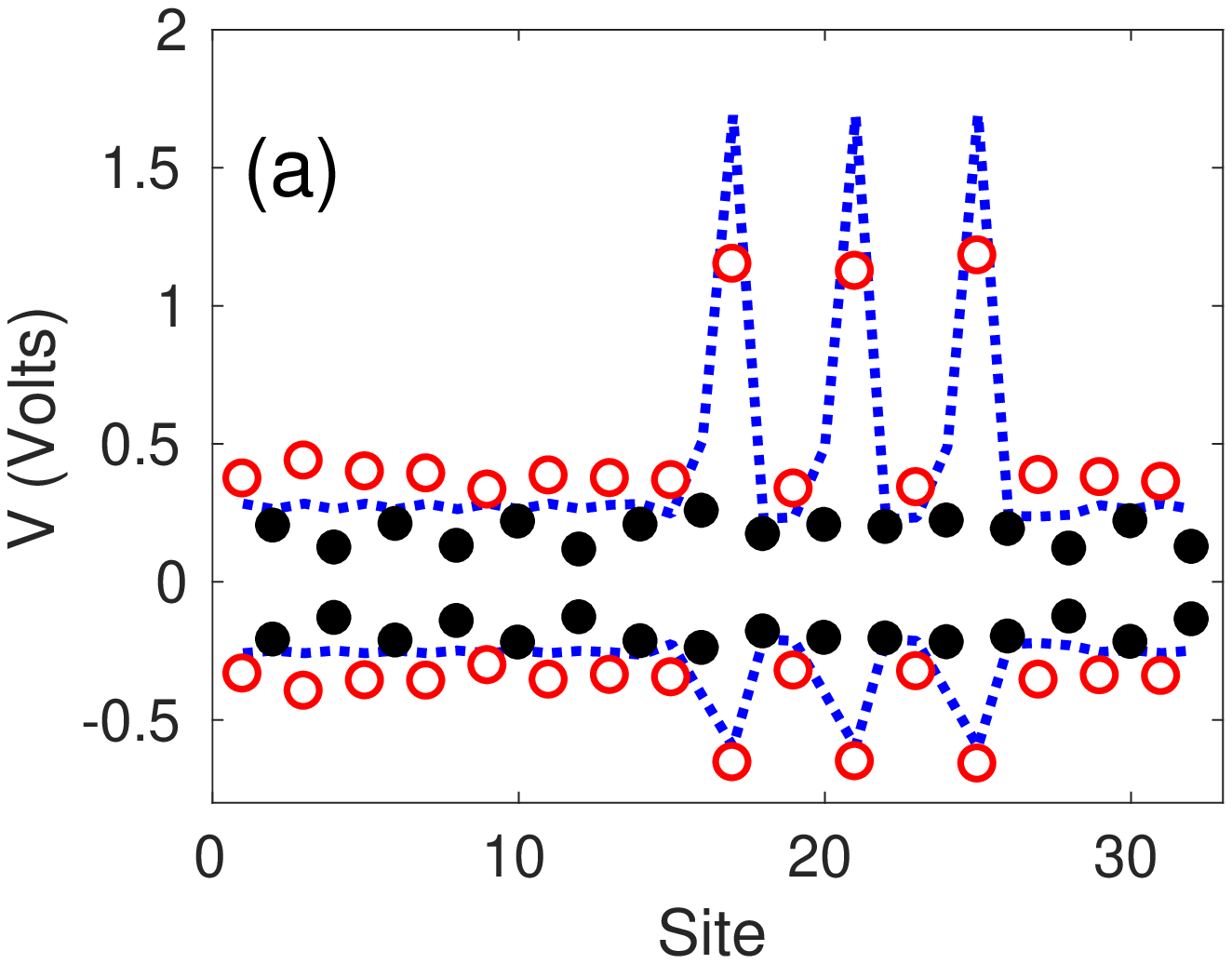}
\includegraphics[width=0.20\textwidth]{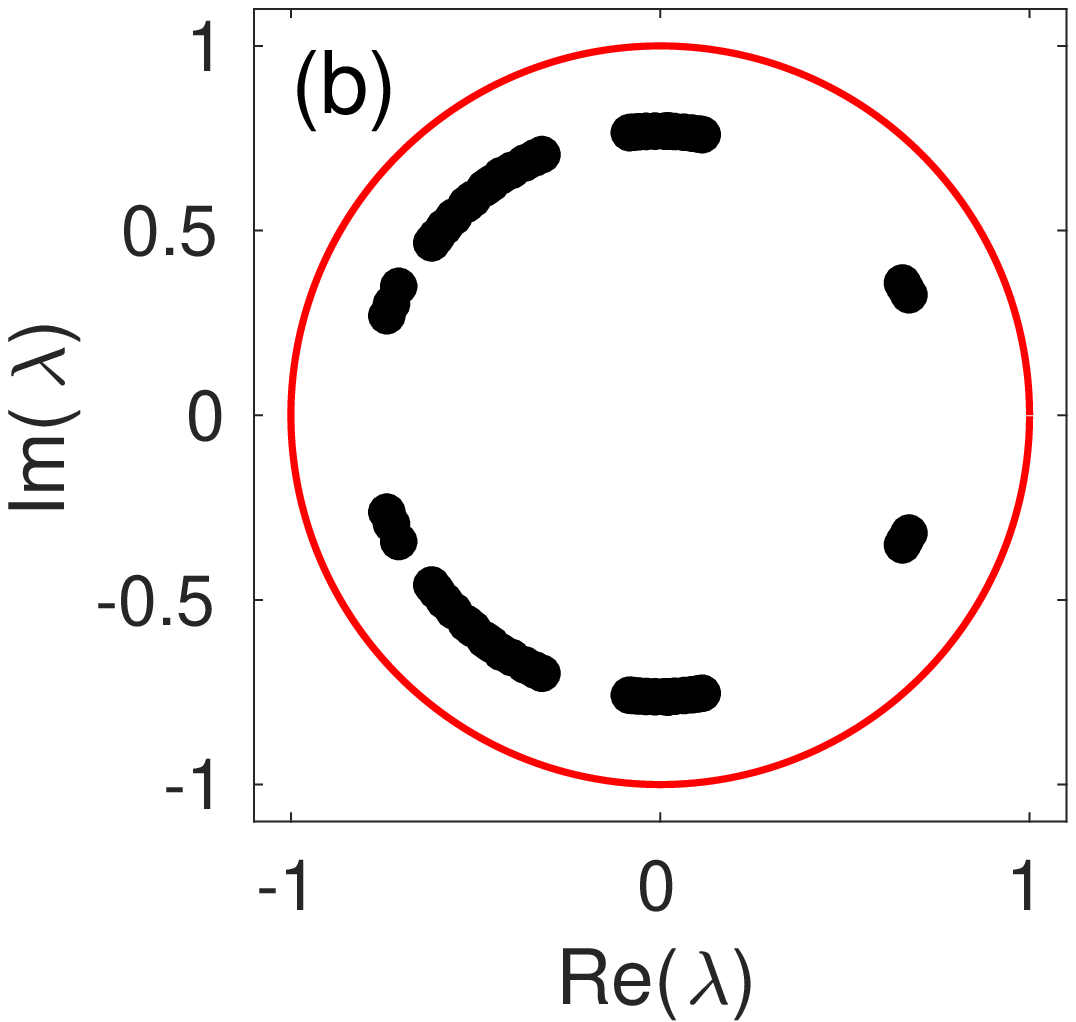}
\caption{(a) Numerical (dotted line) and experimental (circles) three--peak bright breather profile (maximum and minimum amplitude) corresponding to $V_d=3.5$ Volts and $f=532$ kHz. Blank points represent circuit cells of type $(a)$ and black points represent circuit cells of type $(b)$. (b) Floquet multiplier numerical linearization spectrum confirming the stability of the solution.}%
\label{bright_1}	
\end{figure}

\begin{figure}[h]
	\includegraphics[width=0.27\textwidth]{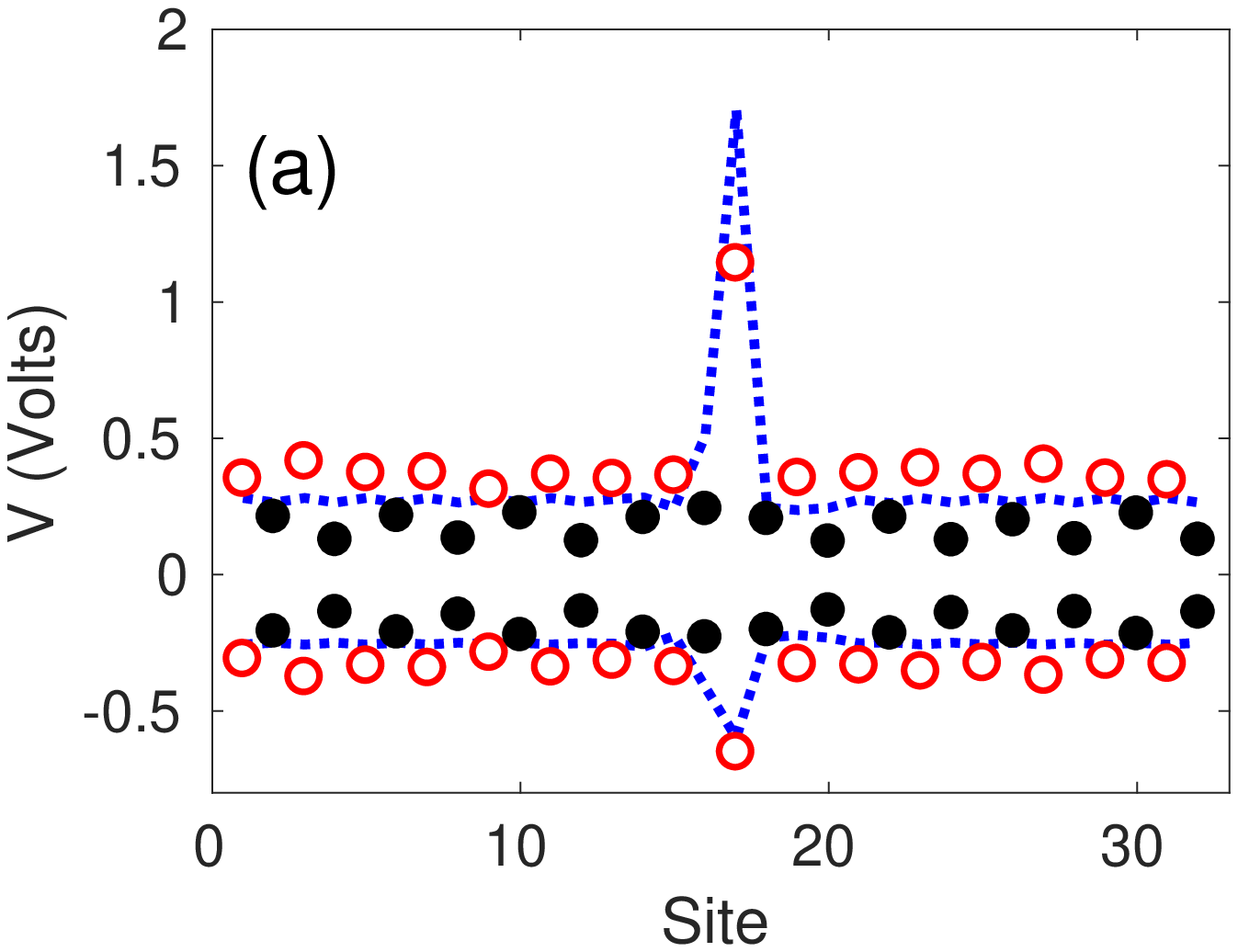}
	\includegraphics[width=0.20\textwidth]{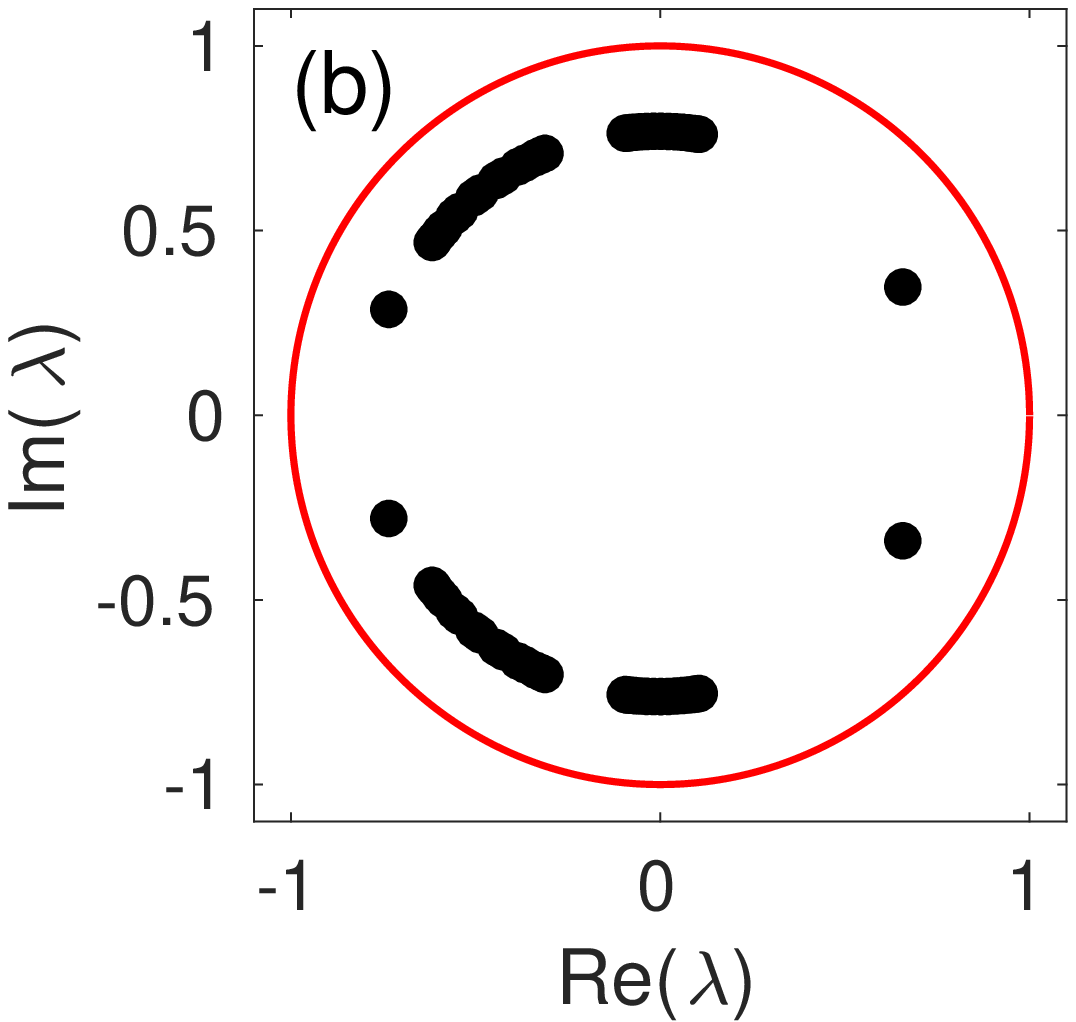}
	\caption{(a) Numerical (dotted line) and experimental (circles) one--peak bright breather profile (maximum and minimum amplitude) corresponding to $V_d=3.5$ Volts and $f=528$ kHz. Blank points represent circuit cells of type $(a)$ and black points represent circuit cells of type $(b)$. (b) Floquet multiplier numerical linearization spectrum confirming, similarly as above, the stability of the solution.}%
	\label{bright_2}
\end{figure}

\begin{figure}[t]
%\begin{center}
\includegraphics[width=0.45\textwidth]{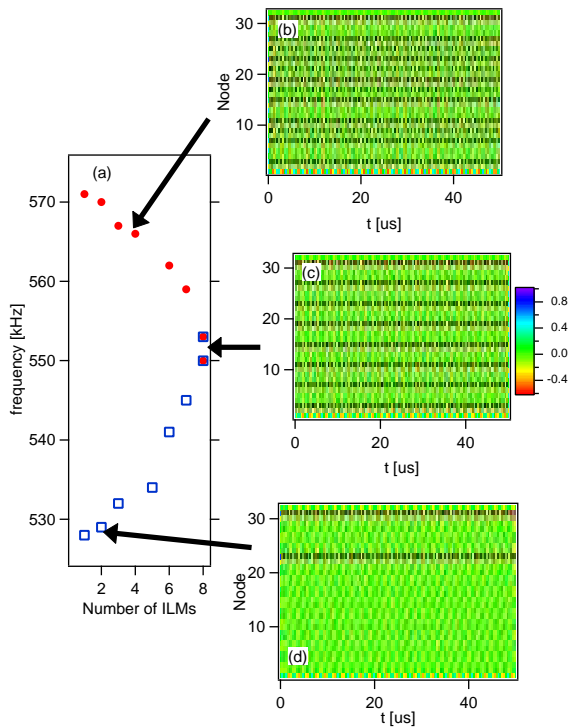}
%\vspace{0.3in}
\caption{Summary of the progression from dark to bright multi-peak ILM structures. Panels (a) show the driver frequency as a function of the number of breathers, both dark (circles, red) and bright (squares, blue). At around 550 kHz, we observe an equal mixture of bright and dark breathers. (b) - (d) show the experimental data corresponding to select frequencies.}
\label{sum1}
%\end{center}
\end{figure}

As we have seen, upon lowering the frequency, the DBs
persist and apparently the number of their stable dips
is increased within the chain.
It is interesting that this is analogous to the phenomenology
observed in the granular case, as discussed in~\cite{granular2}.
Nevertheless, as the frequency is further reduced, an unprecedented
(to our knowledge) scenario develops. More specifically,
the localized pattern inverts, i.e., an excited set of nodes
appears above the background value. Thus, a train of bright (or
perhaps anti-dark~\cite{kivsharbook}) localized modes appears.
If the frequency is further lowered
within the gap, the peaks within
this multi-peak pattern become more sparse until finally only a single
such bright breather survives. Below this lower threshold, the driver
cannot sustain any lattice modes at the given driving amplitude - linear or nonlinear - until we get into the vicinity of the bottom band.
For an amplitude of 3.5V, the experimental threshold lies at 528 kHz,
well above the upper edge of the bottom band.

Figures \ref{bright_1} and \ref{bright_2} depict these bright
localized modes for two different drive frequencies, separated by 4 kHz. At the higher frequency, three ILMs survive, whereas at 528 kHz, we are left with a single bright breather. Let us briefly examine the spatial structure of this state. Its center always resides on the $L_2^{(a)}$ sublattice, namely on a cell with lower inductance, $L_2$. Further, these breathers are extremely sharp - the nearest neighbors on the $L_2^{(b)}$ sublattice register a slightly larger oscillation amplitude compared to the wings, but beyond that no increase beyond background can be discerned. The background oscillation is characterized by the $L_2^{(b)}$ sublattice almost at rest, undergoing only small oscillations, whereas the $L_2^{(a)}$ sublattice is oscillating out-of-phase. Thus, the ILM wings, or background, is consistent with the top (i.e., optic) zone-boundary mode pattern. These observations are in good agreement with numerical results. The precise width of any stability region is hard to match because there exist regions of different peak numbers and regions of different families with the same number of peaks that all
overlap. While the close matching of the numerical and experimental
pictures clearly validates the existence of these structures, a detailed
bifurcation analysis associated with the frequency variations remains
an intriguing topic for future study.

Figure \ref{sum1}(a) effectively summarizes the experimental picture and thus yields a first clue
to the kind of bifurcation diagram mentioned above.
In particular, it depicts the transition from dark to bright breathers as the frequency is lowered starting closer to the bottom of the {top} mode and
moving progressively deeper into the bandgap. At the highest frequency
considered within the gap, the zone-boundary plane-wave mode is excited
at high amplitude. As the frequency is somewhat lowered, first a single dark breather, and then multi-peak dark DBs, are generated. Then, as the frequency is further lowered into the bandgap, eventually a striped nonlinear pattern is seen (Fig.~\ref{sum1}(c)), which could also be characterized as a (roughly) equal mixture of bright and dark breathers. Below that frequency, we observe bright breathers, whose numbers in the lattice decrease as the drive frequency is further lowered. 
It is worth noting that the bright breathers, when multiple of
them arise, are found to always be in phase.

The supporting panels, Fig.~\ref{sum1}(b)-(d), depict the experimental data in time and space at select drive frequencies.
This is one of the especially appealing aspects of this system, namely
the ability for distributed characterization/visualization
of the dynamics (in addition
to the distributed driving capabilities for this system).
Note that the pattern characterizing equal numbers of bright and dark ILMs is qualitatively different from the nonlinear zone-boundary mode. In fact, its spatial periodicity is doubled, and one could describe the resulting pattern as a dynamically induced superlattice. 
We also note that while the dark breathers are sharply localized in this system (as are the bright ones), they do encompass several lattice nodes. Finally, the dark breathers can be shepherded spatially via temporarily induced impurities at neighboring sites. This tunability can be important for the controllable
transfer of energy within such a system.

\subsection{Near the bottom  band}
\begin{figure}[h]
\includegraphics[width=0.27\textwidth]{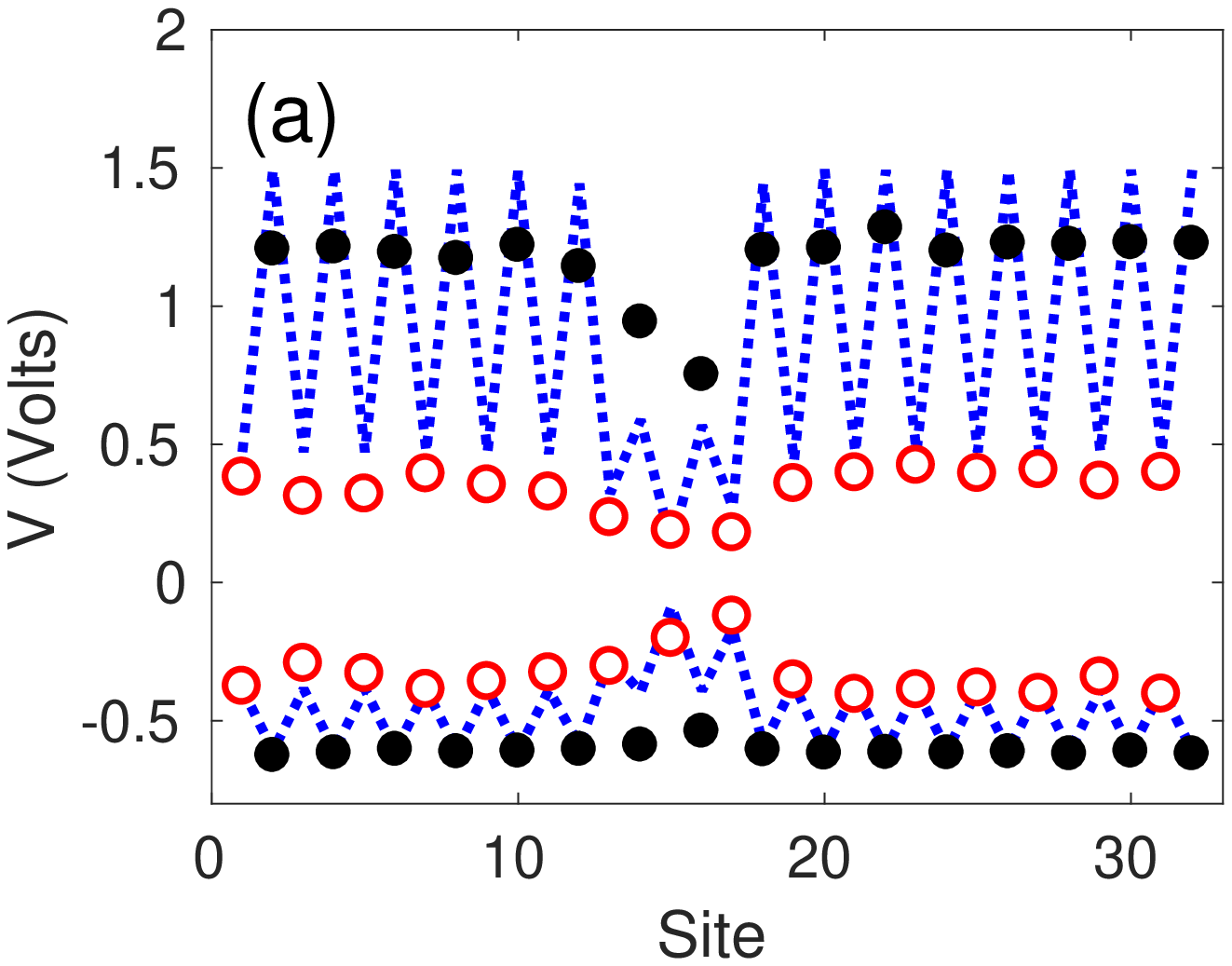}
\includegraphics[width=0.20\textwidth]{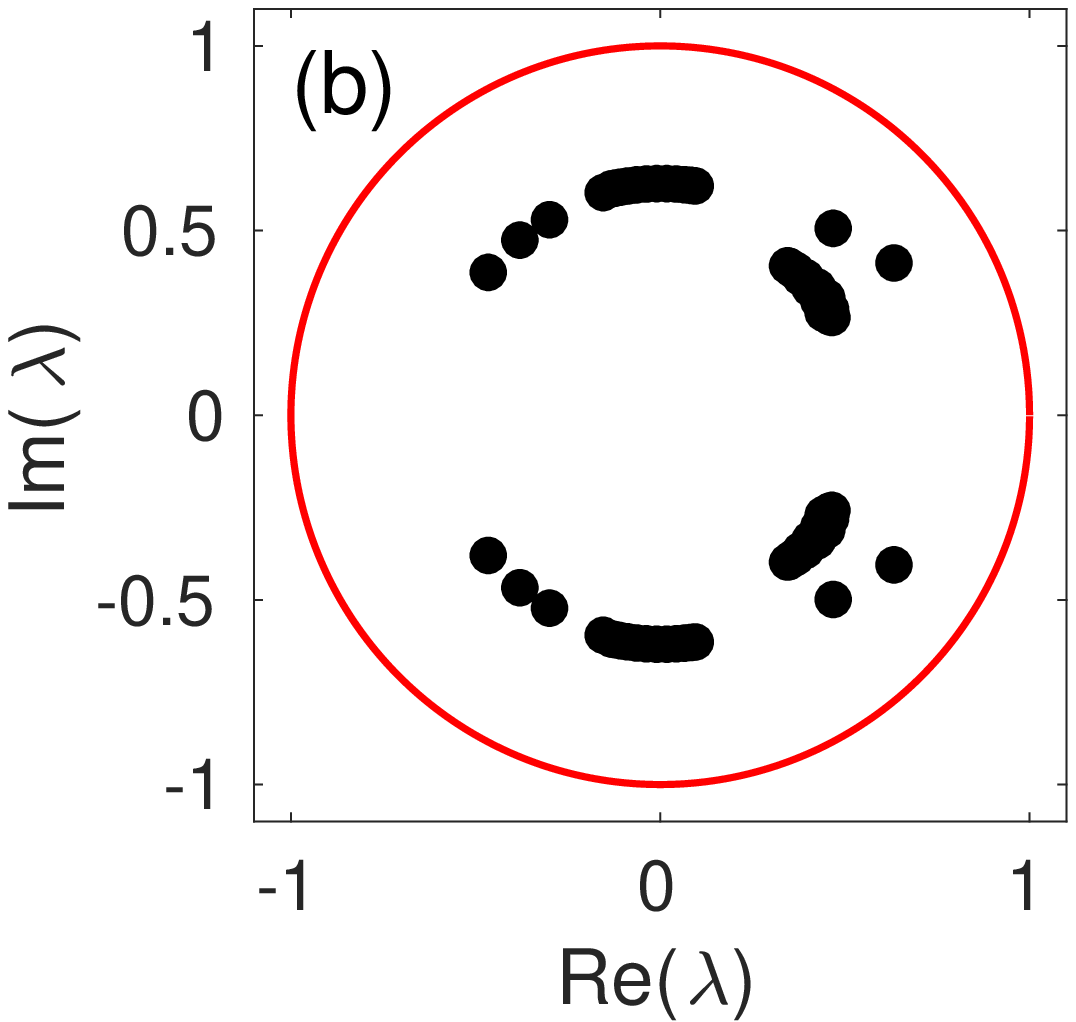}
\caption{(a) Numerical (dotted line) and experimental (circles) dark breather profile (maximum and minimum amplitude) corresponding to $V_d=3.5$ Volts and $f=280$ kHz. Blank points represent circuit cells of type $(a)$ and black points represent circuit cells of type $(b)$. (b) Floquet multiplier numerical linearization spectrum  confirming  the stability of the solution.}%
\label{dark_acous}
\end{figure}

The vicinity of the {bottom} band also features bright and dark ILMs. Figure \ref{dark_acous} depicts the situation at a driver frequency of 280 kHz, just below the bottom of the band. One dark breather is clearly visible at site n=13 in the experimental data. Simulations also find this dark breather, although the background amplitude is slightly higher than in the experiment under identical driving conditions. Floquet analysis again shows that this mode is dynamically stable; see Fig.~\ref{dark_acous}(b).

\begin{figure}[h]
\includegraphics[width=0.48\textwidth]{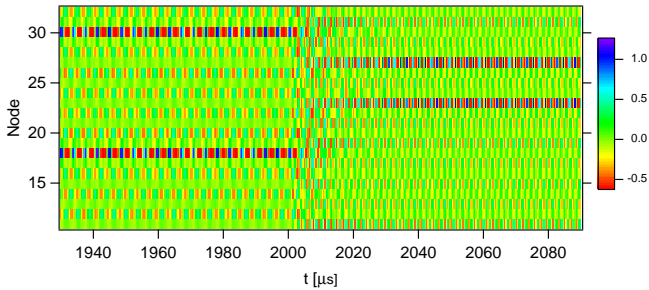}
\caption{Experimental FSK modulation that switches abruptly between two distinct values of the driving frequency while keeping the amplitude fixed at 3.5 V.
  Initially, for $f=$271 kHz, two breathers are seen, but then when the frequency switches to 543 kHz, two distinct ILMs at a different location with
  the latter frequency (centered on odd sites) develop.}%
\label{fsk}
\end{figure}

If the driving frequency is lowered by a mere 10 kHz to 270 kHz, bright ILMs are generated. These bright ILMs just below the {bottom} band are different from the bright ILMs of the {top} band in that their localization centers are on the opposite sublattice, namely the sublattice of $L_2^{(b)}$ sites characterized by the larger inductance value. The situation is succinctly depicted in Fig.~\ref{fsk} where so-called {\it frequency shift key} modulation (FSK) is employed to switch the driver frequency abruptly from one value to another. Here a shift from 271 kHz to 543 kHz was initiated in the experiment at a time of 2000 $\mu$s. The first frequency is consistent with two bright breathers with the associated (lower) frequency,
whereas the second generates two bright breathers with the
corresponding (higher) frequency in a different location.
We see that the breathers are initially located on even sites, but after 2000 $\mu$s quickly settle on odd sites.

\section{Conclusions \& Future Work}

In the present work, we have examined the setting of electrical
lattices in the form of a controllable/tunable diatomic-like system. Building on the
modeling successfully used earlier for the monatomic case, we have
considered both the linear and especially the nonlinear properties
of this system, examining them in parallel both at the experimental and numerical
level. We have found these lattices to be very useful playgrounds for
the exploration of dark breather and multi-breather structures. Such
states have spontaneously emerged in the experiments at different frequencies
within the bandgap near the top band. However, as one further lowers the frequency, a bright nonlinear structure (on top of a
background) emerged, involving initially multiple peaks, which then become progressively
fewer (as the frequency was further lowered), before eventually disappearing altogether
well above the cutoff frequency of the bottom band.

While the structures reported here, we feel, are intriguing, they also
pose numerous open questions for future consideration. Many of these
can be addressed by a systematic analysis of the bifurcations of
single and multiple dark and bright breathers in the bandgap.
Such an analysis, while computationally demanding, would offer a more
systematic roadmap for future experiments. There we could explore
what happens in the immediate vicinity of the (upper edge of the)
bottom branch, as well as the (lower edge of the) top branch. 
Extensions of these fairly controllable structures would also be
worthwhile to consider in higher dimensional systems. Some of these
questions are currently under consideration and will be reported
in future publications. 

\begin{acknowledgments}
F. Palmero visited Dickinson College with support from the VI Plan Propio of the University of Seville. X.-L. Chen was also at Dickinson College for part of this study. This material is based upon work supported by the 
National Science Foundation under Grant No. PHY-1602994 and 
Grant No. DMS-1809074 and the US-AFOSR via FA9550-17-1-0114. 
\end{acknowledgments}

\end{document}